\documentclass[aps,pra,twocolumn]{revtex4}
\usepackage{amsmath}
\usepackage{amscd}
\usepackage{graphicx}
\usepackage{subfigure}
\usepackage{appendix}
\usepackage{amsmath}

\begin{document}
\title[Short Title]{Quantum state transfer in spin chains via shortcuts to adiabaticity }

\author{Bi-Hua Huang$^{1,2}$}
\author{Yi-Hao Kang$^{1,2}$}
\author{Ye-Hong Chen$^{1,2}$}
\author{Zhi-Cheng Shi$^{1,2}$}
\author{Jie Song$^{3}$}
\author{Yan Xia$^{1,2,}$\footnote{E-mail: xia-208@163.com}}

\affiliation{$^{1}$Department of Physics, Fuzhou University, Fuzhou 350116, China\\
             $^{2}$Fujian Key Laboratory of Quantum Information and Quantum Optics (Fuzhou University), Fuzhou 350116, China\\
             $^{3}$Department of Physics, Harbin Institute of Technology, Harbin 150001, China}

\begin{abstract}
Based on shortcuts to adiabaticity and quantum Zeno dynamics, we present a protocol to implement quantum state transfer (QST) in a quantum spin-1/2 chain. In the protocol, the complex Hamiltonian of an $N$-site system is simplified, and a simple effective Hamiltonian is present. It is shown that only the control of the coupling strengths between the boundary spins and the bulk spins are required for QST. Numerical simulations demonstrate that the protocol possesses high efficiency and is robust against the decay and the fluctuations of the control fields. The protocol might provide an alternative choice for transferring quantum states via spin chain systems.

\end{abstract}

\maketitle

\section{Introduction}

Reliable quantum state transfer (QST) between distant locations is one of the pivotal tasks in quantum information processing. To facilitate the QST, an efficient quantum communication channel is essential. There are many systems that can be quantum channel candidates for QST, such as
phonons in ion traps \cite{LeibfriedNAT03422,SchmidtNAT03408}, electrons in semiconductors \cite{GreentreePRB0470},
flux qubits in superconductors \cite{SillanpaaNAT07438, YouNAT11474, MajerNAT07449}, photons in optics
\cite{BlinovNAT04428, DuanQIP0404, MoehringNAT07449, ToganNAT10466}, etc.. Recently, the
spin chain systems have also been envisioned to be good candidates for such channels \cite{BosePRL0391}.
In fact, because the Hamiltonian equivalent to that of a spin chain may be realized in a wide physical systems (e.g., arrays of quantum dots \cite{PetrosyanOC06264}, arrays of Josephson junctions \cite{TsomokosNJP0709,RomitoPRB0571}, cold atoms in optical lattices \cite{DuanPRL0391}, etc.), interests in spin chains continue to increase and many researches based on spin chains have been carried out.

Actually, in the past decades, protocols for efficient QST in spin chains were developed. Researchers found that
whether quantum
information could be propagated perfectly from one end of the chain to the opposite end depended on the distribution of the coupling strengths between the spins \cite{BosePRL0391,ChristandlPRA0571,KarbachPRA0572,KayPRA0673,ChristandlPRL0492}.
For example, state transfer cannot be achieved with perfect fidelity in a uniform Heisenberg spin chain with length $N\geq4$ \cite{BosePRL0391,ChristandlPRA0571}, although it is possible when the couplings are nonuniform and can be individually engineered \cite{ChristandlPRA0571,KarbachPRA0572,KayPRA0673,ChristandlPRL0492,AlbanesePRL0493,FrancoPRL08101,ZwickPRA1184,WuPRA0980,ShiPRA1591}.
It is also possible to
assure high-fidelity QST by appropriately switching the
couplings \cite{EckertNJP0709} in order to induce adiabatic transfer by applying
global \cite{FitzsimonsPRL0697} or local \cite{BalachandranPRA0877, MurphyPRA1082} external fields.
However, these proposes may potentially introduce challenging in practise, because achieving specific control over the whole spin chain
may be quite demanding for long chains which possess a large number of control parameters, especially, when the number of spins grows, the sensitivity to imperfections may greatly increase.
Therefore, protocols involving control only the two boundary couplings rather than
the whole chain seem more feasible.
Such as, one may use the uniform intermediate spins as the interaction mediator (spin bus), then under the condition the boundary spins are weakly-coupled to the bus, quantum states can be transmitted with arbitrary high fidelity \cite{WojcikPRA0572, VenutiPRL0799, VenutiPRA0776, GiampaoloNJP1012,OhPRA1184,Banchi11106,Korzekwa1489}. A usual drawback of this kind mechanism is that they typically require long QST times.
Yet, increasing the transmission times may spoil the process because of the ubiquitous decoherence. High-fidelity and high-speed transmissions seem to be in a dilemma.

In this respect, we note that, recently, a technique named ``shortcuts to adiabaticity" (STA) \cite{MugaAAMOP1362,Demirplak0308,CampoSR1202,ChenXPRL10105,BerryJPA0942,MugaJPB0942,MasudaPRA1184}, which aims at optimally designing Hamiltonian to speed up the quantum adiabatic process, has been put forward.
 The core of the STA is to drive the system following a relatively rapid adiabatic-like process which is not really adiabatic but
leading to the same goals as the adiabatic process does.
 With several advantages, the STA has
attracted a lot of interests \cite{IbanezPRL12109,IbanezPRL1387,AdelPRL13111,BaksicPRL16116,ChenYHPRA1795,
CampoEPL1196,Torrontegui1183,MugaJPB1043,DeffnerPRX1404} and been applied in fields including fast population transfer \cite{ChenYHPRA1489,HuangPRA1796,ZhouNP1713}, fast entanglement generation \cite{KangPRA1694,LuMPRA1489,HuangBHLPL16}, fast quantum computation \cite{ChenYHPRA1591}, and so on \cite{TorosovPRA1489,VacantiNJP1416,SalaPRA1694,DeffnerNJP1618,SongPRA1795,OpatrnyNJP1416}. Experimental implementations of this technique have also been reported \cite{DuNC1607, AnNC1607, BasonNP1208,ZhouNP1713}.
Lately, the idea of STA has been employed to assist and speed-up the adiabatic dynamics when crossing a quantum phase transition in multi-body systems \cite{CampoPRL12109,SaberiPRA1490}. Despite such potential,
the complexity of the Hamiltonian in the multi-body system brings troubles to the application of STA in this field.

On the other hand, quantum Zeno effect (QZE) \cite{MisraJMP7718, CookPST8821, Itano9041} is an interesting phenomenon in quantum mechanics. This effect argues that
frequent measurements can inhibit the transitions between quantum states. In 2002, Facchi \textit{et al}. \cite{FacchiPRL0289} extended the concept of QZE and suggested that the evolution of the system did not necessarily be hindered and could remain in the Zeno subspace defined by the measurements as long as the measurements could be devised with
multidimensional projections.
  This is called quantum Zeno dynamics (QZD) \cite{FacchiJP09196, RaimondPRL10105, RaimondPRA1286, SignolesNP1410}, which has also been successfully observed in a variety of
systems \cite{HostenNAT06439, NagelsPRL9779, KwiatPRL9983, BernuPRL08101}. In fact, QZD can be achieved via
continuous coupling between the system and an external
system instead of discontinuous measurements. Suppose that a dynamical evolution of a system can be governed by the Hamiltonian $H_K=H_{obs}+KH_{meas}$, where $H_{obs}$ is the Hamiltonian of the quantum system, $H_{meas}$ is an interaction Hamiltonian performing the measurement, and $K$ is a coupling constant. In the strong coupling limit $K\rightarrow\infty$, the whole system is governed by the effective Hamiltonian (also known as the ``Zeno Hamiltonian") $H_Z=\sum_{\xi}(\lambda_{\xi}P_{\xi}+P_{\xi}H_{obs}P_{\xi})$, where $P_{\xi}$ is the eigenprojectors of
$H_{meas}$ with eigenvalues $\lambda_{\xi}$, i.e., $H_{meas}=\sum_{\xi}\lambda_{\xi}P_{\xi}$.

Inspired by the technique of QZD, and considering the advantages of STA, in the paper, based on combining use of the two techniques, we present an alternative protocol to implement QST with high fidelity in a one-dimensional quantum spin-1/2 chain. We show that, with the help of QZD,
the complex Hamiltonian of the multi-body system can be simplified and QST can be successfully implemented by merely controlling the couplings between
the boundary spins and the bulk spins. To show the effectiveness of the protocol, we numerically investigate the performance of the protocol when external perturbations exist. The result shows that the protocol holds good robustness against parameter
deviations and dissipations.

The paper is organized as follows. In Sec.~\ref{section:II}, we describe the spin chain system used for quantum state transfer in detail and simply the complex Hamiltonian of a multi-body system into a simple one. In Sec.~\ref{section:III}, we review the shortcut method present in Ref.~\cite{HuangPRA1796} and
 apply the method to inverse design the control Hamiltonian for QST. In Sec.~\ref{section:IV}, we study the effectiveness and the robustness of the protocol.
  Finally, in Sec.~\ref{section:V}, we draw the conclusions.

\section{THEORETICAL MODEL}\label{section:II}

\begin{figure}
 \scalebox{0.2}{\includegraphics{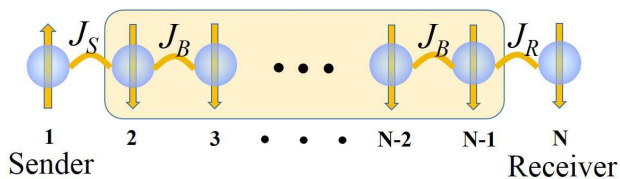}}
  \caption{
         Schematic of the spin chain system. The goal is to transfer the state of the sender qubit to the receiver qubit.
          }
 \label{fig1}
\end{figure}
The system we consider is sketched in Fig.~1. A sender qubit (labelled as $1$) and a receiver qubit (labelled as $N$) are coupled to a uniform spin-1/2 chain at each end. The spin chain in which spins are labelled from $2$ to $N-1$ can be viewed as a spin bus. The Hamiltionian of the system is given as $(\hbar=1)$
 \begin{widetext}
  \begin{eqnarray}\label{eqb1}
     H_N(t)= J_S(t)\overrightarrow{\sigma}_1\cdot\overrightarrow{\sigma}_{2}+J_B\sum_{j=2}^{N-2}\overrightarrow{\sigma}_j\cdot\overrightarrow{\sigma}_{j+1}+J_R(t)\overrightarrow{\sigma}_{N-1}\cdot\overrightarrow{\sigma}_{N},
  \end{eqnarray}
   \end{widetext}
where $\overrightarrow{\sigma}=(\sigma_x,\sigma_y,\sigma_z)$ are the Pauli spin operators, $J_B$ is the uniform nearest-neighbor coupling of the bus spins, and $J_S(t),J_R(t)$ are time-dependent interaction couplings to the spin bus.
In this work, we assume that the qubit-bus couplings are weak, namely, $J_S(t),J_R(t)\ll J_B$.
Clearly,
$[H_N(t),\sum_{n=1}^{N}\sigma_n^z]=0$, which means the number of excited spins is conserved.

For brevity, we quantize the state of each spin as $|0\rangle=|\downarrow\rangle$ (spin down with respect to $z$) and $|1\rangle=|\uparrow\rangle$ (spin up), then the vacuum state is  $|\textbf{0}\rangle=|00\cdots 0\rangle$ and the one-excite state $|\textbf{n}\rangle$ are
\begin{eqnarray}\label{eqb2}
  |\textbf{n}\rangle=\bigotimes_{m=1}^{N}|\delta_{mn}\rangle,~~n=1,2,\cdots N,
  \end{eqnarray}
where $\delta_{mn}$ is the Kronecker delta.
Usually, the aim of QST is to transmit a quantum state $|\psi(0)\rangle=a|\textbf{0}\rangle+b|\textbf{1}\rangle$ encoded on the sender
spin to the receiver spin, where $|a|^2+|b|^2=1$. That is, after a fixed evolution time $t=T$, the system should evolve into the state $|\psi(T)\rangle=a|\textbf{0}\rangle+b|\textbf{N}\rangle$. Since the Hamiltonian $H_N$ conserves the total number of excitations, $|\psi(0)\rangle$ is restricted to evolve within the zero- and one-excitation subspaces. It is obvious that state $|\textbf{0}\rangle$ is conserved during the evolution while $|\textbf{1}\rangle$ will evolves within the subspace spanned by the basis states $\{|\textbf{n}\rangle\}$.

To evaluate how well the channel quality independently
of the specific input state, the transmission fidelity should average
over all possible values of $a$ and
$b$. This leads to quantify the state transfer fidelity with \cite{BosePRL0391}
\begin{eqnarray}\label{eqb3}
     F(t)=\frac{1}{2}+\frac{|f(t)|}{3}+\frac{|f(t)|^2}{6},
  \end{eqnarray}
where $f(t)=\langle \textbf{N}|e^{-iH_Nt}|\textbf{1}\rangle$ is the excitation transition amplitude from the first to the last
spin. Note that $|f(T)|=1$ entails $F(T)=1$ (perfect QST).
In the
following, with the term fidelity we refer to the quantity
given by Eq.~(3).

For more specific, we now focus on the evolution
 within the zero- and one-excitation subspace (denotes as $\forall$) spanned by $N+1$ basis states $\{|\textbf{0}\rangle,|\textbf{n}\rangle\}$. In this subspace, the Hamiltonian $H_N$ can be written in a matrix form as
\begin{eqnarray}\label{eqb4}
  H_N(t)=H_0(t)+H_B,
  \end{eqnarray}
  \begin{widetext}
\begin{eqnarray}
H_0(t)\!=\!\left[%
\begin{array}{cccccccccccccccc}
  J_R+J_S &\  0   &\  0   &\ 0   &\ 0   &\ \cdots &\ 0   &\ 0 &\ 0   &\ 0   \\
  0   &\ J_R-J_S &\  2J_S   &\  0   &\ 0   &\  \cdots &\ 0   &\ 0 &\ 0   &\ 0   \\
  0   &\ 2J_S   &\  J_R-J_S &\  0   &\ 0   &\  \cdots &\ 0   &\ 0 &\ 0   &\ 0   \\
  0   &\ 0   &\  0   &\  J_R+J_S &\ 0   &\  \cdots &\ 0   &\ 0 &\ 0   &\ 0   \\
  0   &\ 0   &\  0   &\  0   &\ J_R+J_S &\  \cdots &\ 0   &\ 0 &\ 0   &\ 0   \\
  \vdots &\ \vdots &\ \vdots &\ \vdots &\ \vdots &\  \ddots &\ \vdots &\ \vdots &\ \vdots &\ \vdots \\
  0   &\ 0   &\  0   &\  0   &\ 0   &\  \cdots &\ J_R+J_S &\ 0 &\ 0   &\ 0   \\
  0   &\ 0   &\  0   &\  0   &\ 0   &\  \cdots &\ 0   &\ J_R+J_S &\ 0 &\ 0   \\
  0   &\ 0   &\  0   &\  0   &\ 0   &\  \cdots &\ 0   &\ 0 &\ J_S-J_R &\ 2J_R   \\
  0   &\ 0   &\  0   &\  0   &\ 0   &\  \cdots &\ 0   &\ 0 &\ 2J_R   &\ J_S-J_R \\
\end{array}%
\right]_{(N+1)\times (N+1)},
\end{eqnarray}
\end{widetext}
\begin{widetext}
\begin{eqnarray}
H_B=J_B\left[%
\begin{array}{cccccccccccccc}
  N-3 &\  0   &\  0   &\ 0   &\ 0   &\  \cdots &\ 0   &\ 0 &\ 0   &\ 0  \\
  0   &\ N-3 &\  0   &\  0   &\ 0   &\  \cdots &\ 0   &\ 0 &\ 0   &\ 0   \\
  0   &\ 0   &\  N-5 &\  2   &\ 0   &\  \cdots &\ 0   &\ 0 &\ 0   &\ 0   \\
  0   &\ 0   &\  2   &\  N-7 &\ 2   &\  \cdots &\ 0   &\ 0 &\ 0   &\ 0   \\
  0   &\ 0   &\  0   &\  2   &\ N-7 &\  \cdots &\ 0   &\ 0 &\ 0   &\ 0   \\
  \vdots &\ \vdots &\ \vdots &\ \vdots &\ \vdots &\ \ddots &\ \vdots &\ \vdots &\ \vdots &\ \vdots \\
  0   &\ 0   &\  0   &\  0   &\ 0   &\ \cdots &\ N-7 &\ 2 &\ 0   &\ 0   \\
  0   &\ 0   &\  0   &\  0   &\ 0   &\  \cdots &\ 2   &\ N-7 &\ 2 &\ 0   \\
  0   &\ 0   &\  0   &\  0   &\ 0   &\  \cdots &\ 0   &\ 2 &\ N-5 &\ 0   \\
  0   &\ 0   &\  0   &\  0   &\ 0   &\  \cdots &\ 0   &\ 0 &\ 0   &\ N-3 \\
\end{array}%
\right]_{(N+1)\times (N+1)}.
\end{eqnarray}
\end{widetext}
It is obvious that $|\textbf{0}\rangle$, $|\textbf{1}\rangle$ and $|\textbf{N}\rangle$ are all
the eigenstates of $H_B$ with the same eigenvalue $(N-3)J_B$. Supposing the eigenstates of $H_B$ are $\{|\phi_k\rangle,k=0,1\cdot\cdot\cdot N\}$, we denote $|\phi_0\rangle=|\textbf{0}\rangle$, $|\phi_1\rangle=|\textbf{1}\rangle$ and $|\phi_2\rangle=|\textbf{N}\rangle$, respectively. The other eigenvalues of $H_B$ can be obtained by the $(N-2)\times (N-2)$ matrix
$H_B^{\prime}=2J_B\cdot M+(N-7)J_B\cdot I$, where $I$ denotes the $(N-2)\times (N-2)$ unit matrix and $M$ is a mirror-symmetric matrix as
 \begin{eqnarray}
M&=&\left[%
\begin{array}{cccccccccc}
    1 &\  1   &\ 0   &\ 0   &\ \cdots &\ 0   &\ 0 &\ 0  &\ 0   \\
    1   &\  0 &\ 1   &\ 0   &\ \cdots &\ 0   &\ 0 &\ 0   &\ 0   \\
    0   &\  1   &\ 0 &\ 1   &\ \cdots &\ 0   &\ 0 &\ 0  &\ 0    \\
\vdots &\ \vdots &\ \vdots &\ \vdots &\ \ddots &\ \vdots &\ \vdots &\ \vdots  &\ \vdots\\
    0   &\  0   &\ 0   &\ 0   &\ \cdots &\ 1&\ 0 &\ 1  &\ 0    \\
    0   &\  0   &\ 0   &\ 0   &\ \cdots &\ 0   &\ 1 &\ 0  &\ 1 \\
    0   &\  0   &\ 0   &\ 0   &\ \cdots &\ 0   &\ 0 &\ 1  &\ 1  \\
  \end{array}%
\right]_{(N-2)\times (N-2)}.
\end{eqnarray}
The eigenvalues of $M$ are nondegenerate \cite{Parlett,CantoniLAA7613} and can be deduced as \cite{banchiNJP1113}
 \begin{eqnarray}
  \varepsilon_p^{M}=2\cos[\frac{(p-1)\pi}{N-2}],~~p=1,2,\cdot\cdot\cdot,N-2.
  \end{eqnarray}
Therefore, according to Eqs.~(6) and (7), the eigenvalues of $H_B^{\prime}$ can be written as
\begin{eqnarray}
  \varepsilon_p^{B}=4J_B\cos[\frac{(p-1)\pi}{N-2}]+(N-7)J_B,
  \end{eqnarray}
 when $p=1$, we get $\varepsilon_1^{B}=(N-3)J_B$, which means that other than the eigenstates $|\phi_0\rangle$, $|\phi_1\rangle$ and $|\phi_2\rangle$, for Hamiltonian $H_B$, there merely exits another eigenstate (denotes as $|\phi_3\rangle$) with eigenvalue $(N-3)J_B$. Solving the equation $H_B|\phi_3\rangle=(N-3)J_B|\phi_3\rangle$,
 we can get the explicit form of $|\phi_3\rangle$ as
 $|\phi_3\rangle=\frac{1}{\sqrt{N-2}}(|\textbf{2}\rangle+|\textbf{3}\rangle+...+|\textbf{{N-1}}\rangle)$.

Therefore, if we regard $H_0(t)$ and $H_B$ as $H_{obs}$ and $KH_{meas}$ in Sec.~I, respectively, when $J_B\gg J_S,J_R$ is satisfied, according to QZD, the Hilbert subspace $\forall$ is split into $(N-2)$ Zeno subspaces due to the degeneracy of eigenvalues of $H_B$ (see Eq.~(9)). Among all these Zeno subspaces, we focus us on the one decided by the eigenvalue $\varepsilon_1^{B}=(N-3)J_B$,
\begin{eqnarray}\label{eqb5}
  Z_1=\{|\textbf{0}\rangle, |\textbf{1}\rangle,|\textbf{N}\rangle, |\phi_3\rangle\}.
    \end{eqnarray}
The reason for that concern is as follows. According to QZD, when the Zeno condition is satisfied, the whole system will approximatively evolve in an invariant subspace consisting of the initial state. While, in the QST process,
 the system is initially in the state $|\psi(0)\rangle=a|\textbf{0}\rangle+b|\textbf{1}\rangle$, that means we can just pay attention to the Zeno subspace who contains states $\{|\textbf{0}\rangle,|\textbf{1}\rangle\}$. And $Z_1$ is the very one that satisfies. Therefore, according to Ref.~\cite{FacchiPRL0289}, the projector in the $Z_1$ Zeno subspace is
\begin{eqnarray}\label{eqb6}
  P_1=\sum_{\alpha}|\alpha\rangle\langle\alpha|~~(|\alpha\rangle\in Z_1),
    \end{eqnarray}
and the effective Hamiltonian of the system can be written as
\begin{eqnarray}\label{eqb7}
  H_{eff,N}(t)&=&\varepsilon_1^{B}P_1+P_1H_0P_1 \cr\cr
        &=&(J_R+J_S)|\textbf{0}\rangle\langle \textbf{0}|+(J_R-J_S)|\textbf{1}\rangle\langle \textbf{1}| \cr\cr
        &+&\frac{N-4}{N-2}(J_R+J_S)|\phi_3\rangle\langle \phi_3|+(J_S-J_R)|\textbf{N}\rangle\langle \textbf{N}| \cr\cr
        &+&[\frac{2J_S}{\sqrt{N-2}}|\textbf{1}\rangle\langle \phi_3|+\frac{2J_R}{\sqrt{N-2}}|\textbf{N}\rangle\langle \phi_3|+H.c.]
        \cr\cr
        &+&(N-3)J_B[|\textbf{0}\rangle\langle \textbf{0}|+|\textbf{1}\rangle\langle \textbf{1}|+|\textbf{N}\rangle\langle \textbf{N}|\cr\cr
        &+&|\phi_3\rangle\langle \phi_3|].
    \end{eqnarray}
Rewriting $H_{eff,N}(t)$ in the matrix form in the basis order $\{|\textbf{0}\rangle, |\textbf{1}\rangle,|\phi_3\rangle,|\textbf{N}\rangle \}$, we obtain
\begin{eqnarray}\label{eqb8}
     H_{eff,N}(t)\!&=\!& \left[
             \begin{array}{cccccc}
            J_R+J_S  &   0   &    0  &  0    \\
            0  &   J_R-J_S & \frac{2J_S}{\sqrt{N-2}} &  0 \\
            0  &   \frac{2J_S}{\sqrt{N-2}} &  \frac{(N-4)(J_R+J_S)}{N-2} & \frac{2J_R}{\sqrt{N-2}}\\
            0  &   0 & \frac{2J_R}{\sqrt{N-2}} &  J_S-J_R \\
             \end{array}
             \right]
              \cr\cr
             &+&(N-3)J_B\cdot I.
  \end{eqnarray}
Since the term $(N-3)J_B\cdot I$ results a global phase in the evolution, in the following discussion, we will safely neglect this term in the Hamiltonian $H_{eff,N}(t)$.

  Note that, for a three-site spin chain, the Hamiltonian of the system is given as
  \begin{eqnarray}
  H_3(t)=J_S(t)\overrightarrow{\sigma}_1\cdot\overrightarrow{\sigma}_{2}+J_R(t)\overrightarrow{\sigma}_2\cdot\overrightarrow{\sigma}_{3}.
  \end{eqnarray}
  Rewriting $H_3(t)$ in the space spanned by $\{|\textbf{0}\rangle,|\textbf{1}\rangle,|\textbf{2}\rangle,|\textbf{3}\rangle\}$, we obtain
  \begin{eqnarray}\label{eqb9}
           H_3(t)\!=\!\left[
             \begin{array}{cccccccccc}
             J_R+J_S  &   0  &   0  &  0  \\
              0   & J_R-J_S & 2J_S       & 0        \\
             0 &  2J_S       & -J_R-J_S & 2J_R         \\
             0 &  0       &2J_R       & J_S-J_R  \\
              \end{array}
             \right].
  \end{eqnarray}
Comparing Eq.~(15) with Eq.~(13), it would be found that Eq.~(13) also applies to $N=3$ case.

 Therefore, for the problem of transmitting quantum state with spin chain, in the condition $J_B\gg J_S(t),J_R(t)$, the complex Hamiltonian of the multi-body system $H_N$  can be reduced to an effective simple one $H_{eff,N}(t)$. The transmission target can be achieved by the effective Hamiltonian $H_{eff,N}(t)~(N\geq3)$.

\section{  REALIZATION OF STATE TRANSFER
BY SHORTCUTS TO ADIABATICITY }\label{section:III}

\subsection{The shortcut method }

 In this section, we will exploit the shortcut method raised in Ref.~\cite{HuangPRA1796} for inversely designing the Hamiltonian to realize QST. Firstly,
we would like to review the method.

Suppose that the Hamiltonian of
 a two-level system possesses the form
 \begin{eqnarray}
 H_0(t)= g_x(t)\sigma_x+g_z(t)\sigma_z,
 \end{eqnarray}
where $g_x(t),g_z(t)$ are arbitrary real functions of
time and $\sigma_x,\sigma_z$ are Pauli operators.
We introduce a picture transformation
$V(t)=e^{-i\theta\sigma_{y}}e^{\frac{i\beta}{2}\sigma_{z}}$, where $\theta$ and $\beta$ are time-dependent parameters. In the picture defined by $V^{\dag}$, the Hamiltonian is
\begin{eqnarray}\label{eqb10}
    H(t)=V^{\dag}H_0(t)V+i(\partial_t{V^{\dag}})V.
  \end{eqnarray}
If we let $H(t)=f_x(t)\sigma_x$, where $f_x(t)$ is a time-dependent function. Substituting Eq.~(16) into Eq.~(17), we deduce
\begin{eqnarray}\label{eqb11}
     g_x(t)&=&\dot{\theta}\cos2\theta\cot\beta-\frac{\dot{\beta}}{2}\sin2\theta,
     \cr
     g_z(t)&=&-\dot{\theta}\sin2\theta\cot\beta-\frac{\dot{\beta}}{2}\cos2\theta,
     \end{eqnarray}
     and
     \begin{eqnarray}
     f_x(t)=\frac{\dot{\theta}}{\sin\beta}.
     \end{eqnarray}
  That is, if the explicit forms of $\theta(t)$ and $\beta(t)$ are decided, according to  Eqs.~(16) and (18), the Hamiltonian $H_0(t)$ that evolves the dynamics is decided.

Supposing the evolution operator in the picture defined by $V^{\dag}$ is $U_V(t)$, it is obvious that $U_V(t)=e^{-i\int H(t^{\prime})dt^{\prime}}=e^{-i\sigma_x\int{f_x(t^{\prime})dt^{\prime}}}$. Then according to picture transformation, in the original picture, the evolution operator of the system can be expressed as
\begin{eqnarray}\label{eqb12}
\textrm{U}_O(t)=V(t)U_V(t)V^{\dag}(0)=V(t)e^{-i\sigma_x\int_{0}^{t}{f_x(t^{\prime})dt^{\prime}}}V^{\dag}(0),\cr
 \end{eqnarray}
which is parameterized with $\theta,\beta$.
    Since the state of the system at any time can be expressed as
    \begin{eqnarray}
     |\Psi(t)\rangle=\textrm{U}_O(t)|\Psi(0)\rangle,
     \end{eqnarray}
where $|\Psi(0)\rangle$ is the initial state of the system. It implies that if the desired dynamics is known, for example, $|\Psi(0)\rangle$ and $|\Psi(T)\rangle$ are known, according to Eqs.~(20-21), the boundary conditions of the parameters $(\theta,\beta)$ are specified.
Then by appropriately setting the functions of $\theta, \beta$ according to the boundary conditions, the explicit form of Hamiltonian $H_0(t)$ is determined. That means  we have inversely designed the Hamiltonian to engineer the desired evolution.

In the case of $V(0)=1$, Eq.~(20) can be explicitly expressed as
\begin{widetext}
\begin{eqnarray}\label{eqb13}
     \textrm{U}_O(t)
              =\left(
               \begin{array}{cccc}
               e^{\frac{i\beta}{2}}\cos\theta\cos\delta_x+ie^{-\frac{i\beta}{2}}\sin\theta\sin\delta_x ~~& -ie^{\frac{i\beta}{2}}\cos\theta\sin\delta_x-e^{-\frac{i\beta}{2}}\sin\theta\cos\delta_x \\
               e^{\frac{i\beta}{2}}\sin\theta\cos\delta_x-ie^{-\frac{i\beta}{2}}\cos\theta\sin\delta_x  ~~&  -ie^{\frac{i\beta}{2}}\sin\theta\sin\delta_x+e^{-\frac{i\beta}{2}}\cos\theta\cos\delta_x  \\
             \end{array}
            \right),
  \end{eqnarray}
  \end{widetext}
where $\delta_x$ is time-dependent, $\delta_x(t)=\int^{t}_0 \frac{\dot{\theta}}{\sin\beta}dt^{\prime}$.

\subsection{Inverse designation of the Hamiltonian for quantum state transfer}

To inverse design the driving Hamiltonian via the shortcut method described in part A, in the first, we need to make a minor ``surgery'' on the Hamiltonian $H_{eff,N}(t)$, and for that, we
introduce a unitary transformation
  \begin{eqnarray}\label{eqb14}
     S_N= \left[
             \begin{array}{ccccc}
              1  &   0   &   0   &   0   \\
              0  & \frac{1}{\sqrt{N}} & \frac{\sqrt{N-2}}{\sqrt{N}} &  \frac{1}{\sqrt{N}} \\
              0  & \frac{1}{\sqrt{2}} &  0 & -\frac{1}{\sqrt{2}}\\
              0   & \frac{\sqrt{N-2}}{\sqrt{2N}} & -\frac{2}{\sqrt{2N}} &  \frac{\sqrt{N-2}}{\sqrt{2N}} \\
             \end{array}
            \right].
  \end{eqnarray}
 Performing $S_N$ on $H_{eff,N}(t)$, we obtain
 \begin{widetext}
  \begin{eqnarray}\label{eqb15}
     H_{e,N}(t)&=&S_NH_{eff,N}(t)S_N^{\dag}
          \cr\cr
          &=&\left[
             \begin{array}{cccccc}
             \frac{(N-1)(J_R+J_S)}{N-2}\textbf{I}  &  \textbf{0}   \\
            \textbf{0} &  \frac{(J_R+J_S)}{N-2}\sigma_z+\sqrt{\frac{N}{N-2}}(J_R-J_S)\sigma_x\\
            \end{array}
            \right]
            -\frac{J_R+J_S}{N-2}\left[
          \begin{array}{cccccc}
               \textbf{I} & \textbf{0}\\
               \textbf{0} &  \textbf{I}\\
             \end{array}
            \right],
  \end{eqnarray}
  \end{widetext}
  where $\sigma_{x(z)}$ are Pauli matrix, $\textbf{0}$ and $\textbf{I}$ denote the 2 by 2 zero matrix and unit matrix, respectively (note that in the following part the symbols $\textbf{0}$ and $\textbf{I}$ represent the same meaning as they are in here).

To make $H_{e,N}(t)$ possess the same form as $H_0(t)$ in Eq.~(16), we introduce three operators:
\begin{eqnarray}\label{eqb16}
     \sigma_q^{\prime}= \left[
             \begin{array}{cccc}
               \textbf{0} & \textbf{0} \\
               \textbf{0} & \sigma_q \\
              \end{array}
            \right],~~q=x,y,z,
  \end{eqnarray}
where $\sigma_q$ are Pauli matrix. It's obvious that $\overrightarrow{\sigma^{\prime}}=(\sigma_x^{\prime},\sigma_y^{\prime},\sigma_z^{\prime})$ are Pauli-like operators since $[\sigma_i^{\prime},\sigma_j^{\prime}]=2i\epsilon_{ijk}\sigma_k^{\prime}$.
Therefore, with $\overrightarrow{\sigma^{\prime}}$, the Hamiltonian $H_{e,N}(t)$ can be rewritten as
\begin{eqnarray}\label{eqb17}
     H_{e,N}(t)&=& H_{e0}(t)+H_{e1}(t),\cr
      H_{e1}(t)&=&\sqrt{\frac{N}{N-2}}(J_R-J_S)\sigma_x^{\prime}+\frac{1}{N-2}(J_R+J_S)\sigma_z^{\prime},\cr\cr
     H_{e0}(t)&=& \frac{(N-1)(J_R+J_S)}{N-2}\left[
             \begin{array}{cccccc}
               \textbf{I} & \textbf{0}\\
               \textbf{0} &  \textbf{0}\\
             \end{array}
            \right]
            \cr
          &-&\frac{J_R+J_S}{N-2}\left[
          \begin{array}{cccccc}
               \textbf{I} & \textbf{0}\\
               \textbf{0} &  \textbf{I}\\
             \end{array}
            \right].
       \end{eqnarray}
  Apparently, $H_{e1}(t)$ holds the same form as $H_0(t)$ in Eq.~(16). If we replace $g_x(t),g_z(t),\sigma_{q}$ in Eq.~(16) with $\sqrt{\frac{N}{N-2}}(J_R-J_S), ~\frac{1}{N-2}(J_R+J_S),~\sigma_q^{\prime}$, respectively, we obtain
\begin{eqnarray}\label{eqb18}
     J_R(t)&=&\frac{N-2}{2}g_z(t)+\frac{\sqrt{N-2}}{2\sqrt{N}}g_x(t),
     \cr
     J_S(t)&=&\frac{N-2}{2}g_z(t)-\frac{\sqrt{N-2}}{2\sqrt{N}}g_x(t),
  \end{eqnarray}
that is, once $g_x(t),g_z(t)$ are determined, the couplings $J_S(t),J_R(t)$ are determined too.

Defining $\varphi_N(t)=\frac{N-1}{N-2}\int_{0}^{t}(J_R+J_S)dt^{\prime}$, according to Eq.~(22) and Eq.~(26), the evolution operator of $H_{e,N}(t)$ can be deduced as
\begin{eqnarray}\label{eqb19}
  U_{e,N}(t)=\left(
             \begin{array}{cccc}
                e^{-i\varphi_N(t)}\textbf{I}& \textbf{0}  \\
                \textbf{0} & \textrm{U}_O(t) \\
              \end{array}
            \right),
       \end{eqnarray}
   where $\textrm{U}_O(t)$ is the matrix that appears in Eq.~(22).
    Note that here we have ignored the global phase.

According to Eqs.~(13) and (23) and picture transformation, to realize $(a|\textbf{0}\rangle+b|\textbf{1}\rangle)\rightarrow (a|\textbf{0}\rangle+b|\textbf{N}\rangle)$ in the original picture, in the picture defined by $S_N$, we need to realize the process from $|\chi_1\rangle=[a,\frac{b}{\sqrt{N}},\frac{b}{\sqrt{2}},\frac{b\sqrt{N-2}}{\sqrt{2N}}]^T$ to $|\chi_2\rangle=[a,\frac{b}{\sqrt{N}},-\frac{b}{\sqrt{2}},\frac{b\sqrt{N-2}}{\sqrt{2N}}]^T$ via the dynamics $H_{e,N}(t)$. Namely, we should design appropriate exchange couplings $J_S,J_R$ to carry out the state transfer $|\chi_2\rangle=U_{e,N}(T)|\chi_1\rangle$ (up to a global phase).
To achieve the goal and make the external driving fields could be smoothly
turned on and turned off, the boundary conditions of Eq.~(28) can be set as
\begin{eqnarray}\label{eqb20}
\theta(0)=\theta(T)=0,~\delta_x(0)=\delta_x(T)=0,\cr
  \beta(0)=0,~\beta(T)=N_{\beta}\pi,~\varphi_N(T)=\frac{\beta(T)}{2}+2f\pi,
    \end{eqnarray}
where $N_{\beta}=1,3,5,\cdot\cdot\cdot$ is an odd number and $f$ is an arbitrary integer.
  For example, if we choose $N_{\beta}=1, f=0$ and substitute them into Eqs.~(22) and (28), we can obtain
 \begin{eqnarray}\label{eqb21}
  U_{e,N}(T)=-i\left(
             \begin{array}{cccc}
               1 & 0 & 0 & 0 \\
                0 & 1& 0 & 0  \\
               0 &0 &  -1 & 0 \\
               0 & 0 &  0 & 1  \\
              \end{array}
            \right),
    \end{eqnarray}
therefore, $U_{e,N}(T,0)|\chi_1\rangle=|\chi_2\rangle$ (up to a global phase) is achieved. The QST process has been successfully implemented.

\section{ NUMERICAL SIMULATION AND DISCUSSION }\label{section:IV}

\begin{figure}
 \scalebox{0.3}{\includegraphics{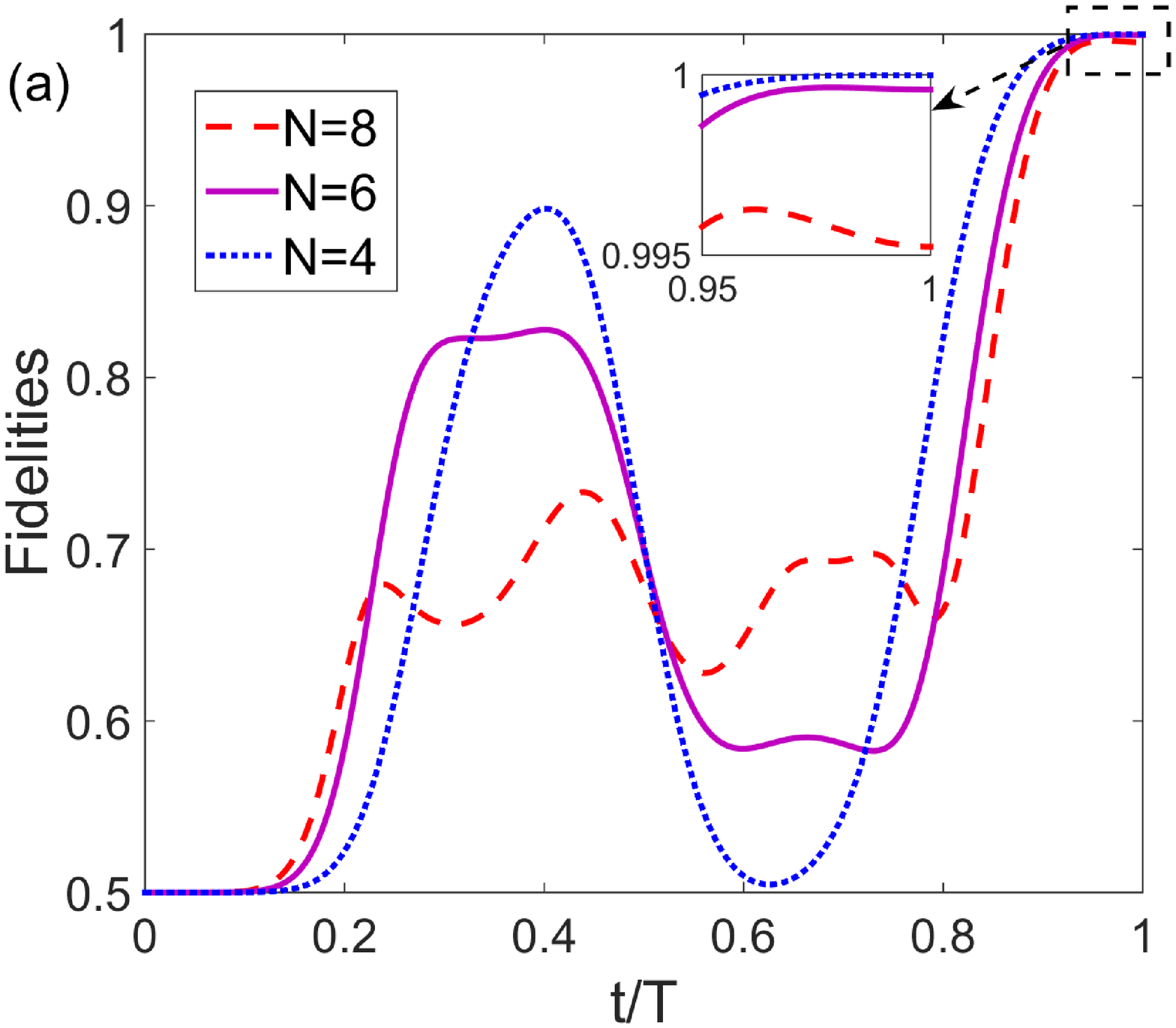}}
 \scalebox{0.3}{\includegraphics{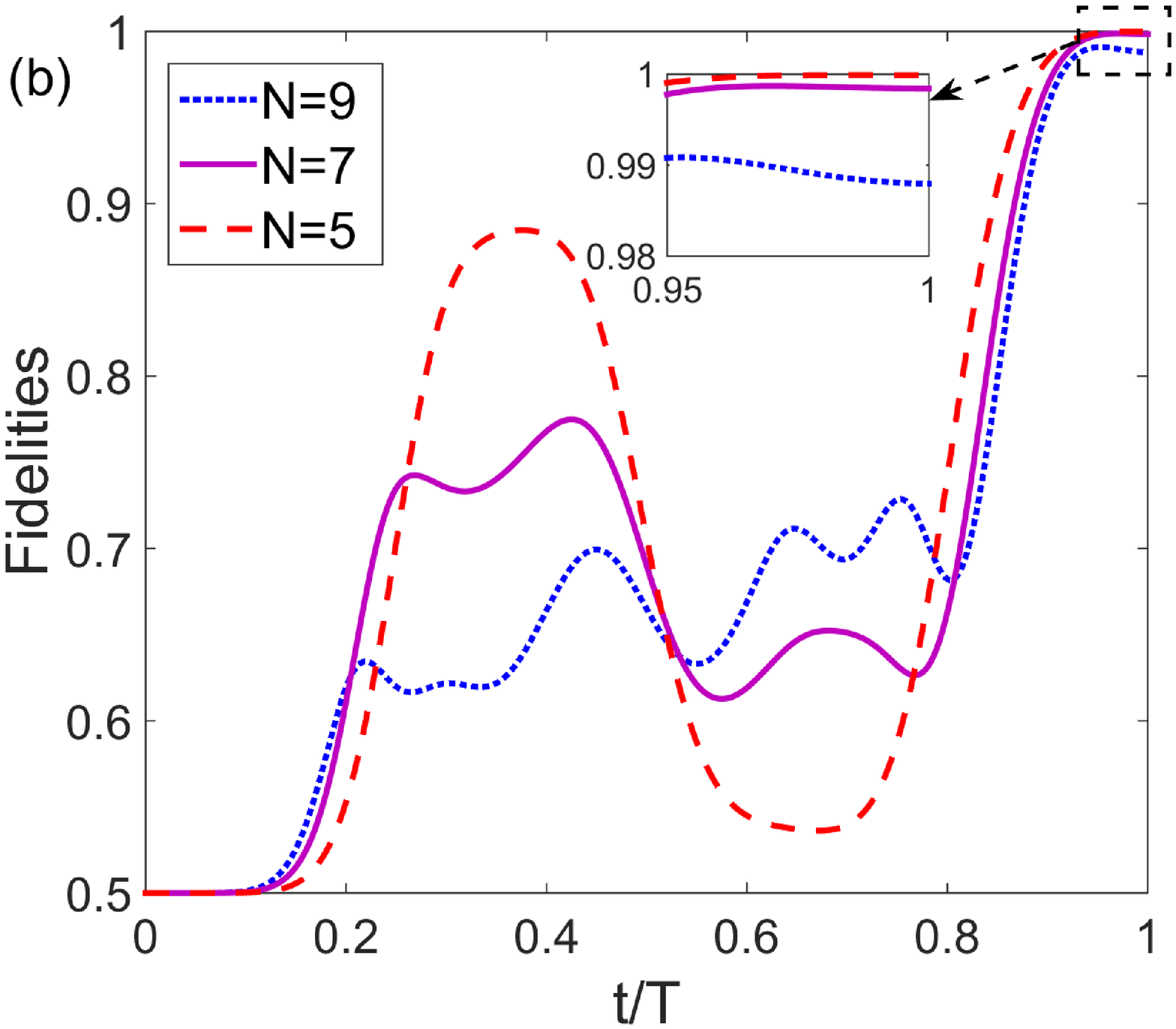}}
  \caption{
         (a) Fidelities versus $t/T$ for spin chain length $N=4,6,8$.
         (b) Fidelities versus $t/T$ for spin chain length $N=5,7,9$.
          }
 \label{fig2}
\end{figure}

\begin{figure}[!htb]
 \scalebox{0.3}{\includegraphics{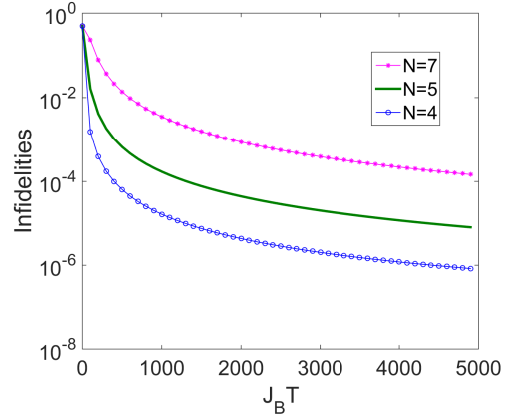}}
   \caption{
         Infideltiies $1-F(T)$ as a function of $J_BT$ with spin chain length $N=4,5,7$.
          }
 \label{fig3}
\end{figure}
\begin{figure}
 \scalebox{0.3}{\includegraphics{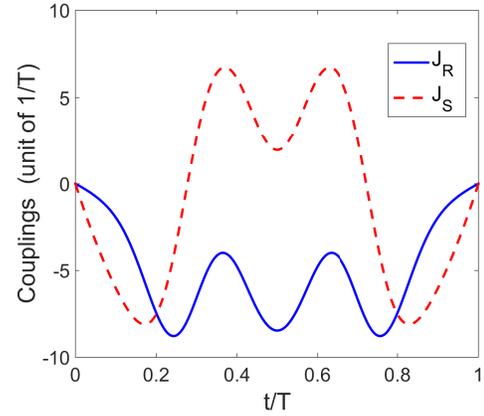}}
   \caption{
         The exchange couplings for $N=5$ case with the parameters set as $\beta(T)=3\pi, \varphi_N(T)=-2.5\pi$.
          }
 \label{fig4}
\end{figure}
In this section, we will numerically investigate the effectiveness of the protocol. First of all, to satisfy the boundary conditions in Eq.~(29), as an example, we select polynomial functions to fit the conditions and the corresponding $\beta(t)$ and $\dot{\delta_x}(t)$ are set as
\begin{eqnarray}\label{eqb22}
     \beta(t)&=&N_{\beta}(\frac{6\pi t^2}{T^2})(\frac{1}{2}-\frac{1}{3}\frac{t}{T}),
     \cr
     \dot{\beta}(t)&=&N_{\beta}\frac{6\pi t}{T^2}(1-\frac{t}{T}),
     \cr
     \dot{\delta_x}(t)&=&\mu(\frac{t}{T^2})(1-\frac{t}{T})(1-2\frac{t}{T}),
  \end{eqnarray}
where $\mu$ should be chosen such that
\begin{eqnarray}\label{eqb23}
    \theta (T)=\int_{0}^{T}\dot{\delta_x}\sin\beta dt=0.
  \end{eqnarray}
  For better comparison, the parameters for different chain length are set with the same values: $\beta(T)=3\pi, \varphi_N(T)=-2.5\pi$. Note that, other choices of the parameters or functions are also available.

According to Eqs.~(1), (18), (27) and (31-32), with $J_B=1000/T$, the fidelities of QST via spin chain with different length
$N=4,5,\cdot\cdot\cdot,9$ are shown in Fig.~2(a) and 2(b), respectively. We use
the fidelity in Eq.~(3) to assess the quality of the QST.
 As seen from Fig.~2, high fidelity can be achieved using the present protocol, although the precision of the final fidelity is slightly different with different length $N$. To see more of the specific details, we draw in Fig.~3 the infidelity $[1-F(T)]$ for different durations of the process and for chains of different length.
As shown in Fig.~3, when the spin length $N$ is fixed, the final fidelity $F(T)$ increases with the increase of $J_BT$. For different length $N$, to achieve the same final fidelity $F(T)$, it is clear that the larger the length $N$, the larger the $J_BT$ required.
The increase of $J_BT$ with growing $N$ can be explained: when $N$ goes up, as seen from Eq.~(27), $J_{R(S)}(t)$ increases accordingly, therefore, to well satisfy the quantum Zeno condition: $J_{R(S)}(t)\ll J_B$, the larger $J_B$ is needed.
 So, the choice of appropriate $J_B$ is related with the chain length $N$ and the required final fidelity. When $N$ is fixed, we can use numerical results to decide how large $J_B$ is sufficient. Here, we take $N=5$ case as an example.
As shown in Fig.~4, for $N=5$ case, a sample of control couplings $J_S(t)$ and $J_R(t)$ is plotted with $\beta(T)=3\pi, \varphi_N(T)=-2.5\pi$. From Fig.~4, the maximum of $J_{R(S)}(t)$ is about $J_M=\max\limits_{0\leq t\leq T}\{J_S(t),J_R(t)\}\sim10/T$. Therefore, if a final fidelity $F(T)$ about 0.99 is required (the infidelity $[1-F(T)]$ is about 0.01), according to Fig.~3, $J_B$ is about $100/T$, which means that $\frac{J_B}{J_M}\sim10$ is sufficient. If the needed final fidelity $F(T)$ is about 0.999 (the infidelity $[1-F(T)]$ is about 0.001), then according to Fig.~3, $\frac{J_B}{J_M}$ is around 40. That is, for fixed $N$, the factor of $\frac{J_B}{J_M}$ depends on the required final fidelity. The higher the final fidelity, the larger the $J_B$ required.

\begin{figure}
 \scalebox{0.3}{\includegraphics{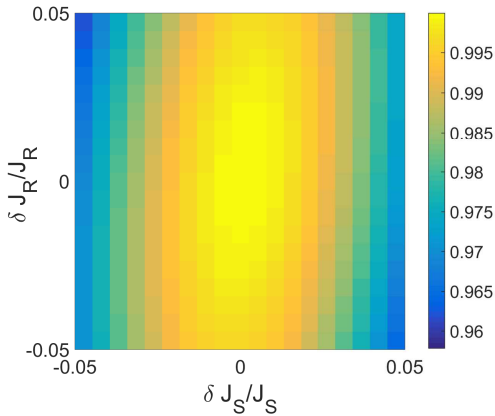}}
   \caption{
         The final fidelity $F(T)$ versus $\delta J_R/J_R$ and $\delta J_S/J_S$ with $J_BT=1000$.
          }
 \label{fig5}
\end{figure}
\begin{figure}
 \scalebox{0.3}{\includegraphics{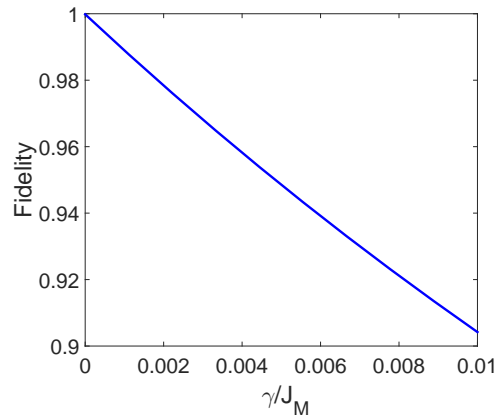}}
   \caption{
         The final fidelity $F(T)$ versus $\gamma/J_M$ with $J_BT=1000$.
          }
 \label{fig6}
\end{figure}

In the following, as an illustrative case, and without loss of generality, we  focus on $N=5$ case and adopt $J_BT=1000$ for more specific discussion because the robustness properties are similar for other value of $N$.
Firstly, in
realistic scenarios, one may expect to have disorder in the
Hamiltonian which deteriorates the quality of the protocol. Here we would like to discuss the disorder effects of the variations $\delta J_S,\delta J_R$ in the couplings $J_S$ and $J_R$, respectively.
Adopting the couplings $J_S(t)$ and $J_R(t)$ shown in Fig.~4 for discussion, we plot $F(T)$ versus $\delta J_S/J_S$ and $\delta J_R/J_R$ in Fig.~5.
As shown in Fig.~5, though the variations $\delta J_S$ and $\delta J_R$ do influence in the fidelity, when $|\delta J_S/J_S|=|\delta J_R/J_R|=5\%$, $F(T)$ is still higher than 0.96. This indicates that the present protocol holds robustness against the disorder in the couplings $J_S, J_R$.

Secondly, in real experiments,
the spin system might not be well isolated from the environment and the dissipations caused by
decay mechanisms are ineluctable. Therefore, checking the fidelity $F(T)$ when decay mechanisms
are taken into account can help us forecast the
experimental feasibility.
Among the multitude of decay mechanisms, dephasing is one of the major challenges \cite{Breuer02}. The original of dephasing is the random energy fluctuations induced on qubit levels by external field. For weak random field fluctuations, the evolution of the system can be described by a master equation \cite{Breuer02,BayatNJP1517}:
\begin{eqnarray}\label{eqb24}
   \dot{\rho}(t)=i[\rho(t),H_N]+\sum_{l=1}^{N}\gamma_l[\sigma_l^z\rho(t)\sigma_l^z-\rho(t)],
  \end{eqnarray}
where the first term in the right hand side is the unitary Schr\"{o}dinger evolution and the second term is the dephasing with rate $\gamma_l$. For simplicity, we assume $\gamma_l=\gamma$ for all the spin particles. As the dephasing is introduced to the spin system, the spins in the system have a chance to jump out of the system.
The final fidelity $F(T)$ versus $\gamma/J_M$ is given in Fig.~6, where $J_M=\max\limits_{0\leq t\leq T}\{J_S(t),J_R(t)\}$ is the maximum of the couplings $J_S(t),J_R(t)$.
According to Fig.~6, the dephasing influences $F(T)$ a bit, when $\gamma/J_M$ increases from 0 to 0.01, $F(T)$ is about 0.91. This implies that the protocol is robust against the dephasing mechanism. Certainly, to reduce the nocuous effects of dephasing, some strategies may be helpful, such as performing regular global measurement on the system \cite{BayatNJP1517}.

\section{CONCLUSION}\label{section:V}

In conclusion, we have developed a hybrid strategy combining shortcuts to adiabaticity and quantum Zeno dynamics that allows us to
achieve
a high-quality quantum state
transfer in a spin chain system. In the protocol,
under the assumption of weak qubit-chain couplings $J_S,J_R\ll J_B$,
the QST can be realized by the controls exerted only on the boundary sites of the chain, which might facilitate the experimental realization.
In addition,
numerical simulations show that the protocol exhibits resilience against
the defasing and operational imperfection.
Note that, besides designing the Hamiltonian via the shortcut method raised in Ref.~\cite{HuangPRA1796}, other shortcut methods \cite{MugaAAMOP1362} are also feasible to perform the task.
Moreover,
the approach might be used across the range of physical hardware types that can be mapped onto the spin chain Hamiltonian. We hope that the protocol might also offer an alternative choice for short distance communications and provide good candidates for the realization of reliable quantum communication in quantum networks.

\section*{ACKNOWLEDGEMENT}

 This work was supported by the National Natural Science Foundation of China under Grants No. 11575045 and No. 11674060, the Major State Basic Research
Development Program of China under Grant No. 2012CB921601, and the
Natural Science Foundation of Fujian Province under Grant No.
JAT160081.

\newpage

\end{document}